# Generalized Short Circuit Ratio for Multi Power Electronic based Devices Infeed Systems: Definition and Theoretical Analysis

Huanhai Xin, *Member IEEE*, Wei Dong, Deqiang Gan, *Senior Member IEEE*, Di Wu, *Member IEEE*, Xiaoming Yuan, *Senior Member IEEE*

*Abstract*—Short circuit ratio (SCR) is widely applied to analyze the strength of AC system and the small signal stability for single power electronic based devices infeed systems (SPEISs). However, there still lacking the theory of short circuit ratio applicable for multi power electronic based devices infeed systems (MPEIS), as the complex coupling among multi power electronic devices (PEDs) leads to difficulties in stability analysis. In this regard, this paper firstly proposes a concept named generalized short circuit ratio (gSCR) to measure the strength of connected AC grid in a multi-infeed system from the small signal stability point of view. Generally, the gSCR is physically and mathematically extended from conventional SCR by decomposing the multi-infeed system into *n* independent single infeed systems. Then the operation gSCR (OgSCR) is proposed based on gSCR in order to take the variation of operation point into consideration. The participation factors and sensitivity are analyzed as well. Finally, simulations are conducted to demonstrate the rationality and effectiveness of the defined gSCR and OgSCR.

## I. Introduction

With fast development of smart grid and urgent demand for flexibility and controllability enhancement of power system, power-electronic-based devices (PEDs) are increasing used for the integration of renewable energy generations, VAR compensation devices in ac transmissions systems [1-3]. As a result, a large number of PEDs will be accessed to the AC power grid. These PEDs and the AC grid together constitute the multi power electronic based devices infeed systems (MPEIS).

The high penetration of PEDs inevitably enlarges the equivalent AC grid impedance and weakens the AC grid, which makes the interactions between PEDs and AC grid more complex [4-5]. As a result, the risk of oscillation issues arises or becomes more critical in a weak AC grid [6-7]. The previous researches show that the characteristics of AC grid play an important role in the stability of grid-connected PEDs system: the structure and parameters of AC grid affect the resonance of PEDs filter, and the strength of AC grid affect the interactions among PEDs. For this reason, to quantitatively evaluate the strength of AC grid and the interactions of PEDs is in urgent demand.

In theoretical research and engineering application, short circuit ratio (SCR) is widely applied to measure the strength of AC grid and evaluate the small signal stability of grid-connected PED system qualitatively [8-10]. Lager SCR leads to a stronger AC grid and a more stable system, and vice versa. SCR is also applied to measure the static voltage stability of AC-DC systems, which involves two indices, i.e., the critical SCR (CSCR) and the boundary SCR (BSCR) [11-12]. Usually, $CSCR \approx 2$ is used to differentiate very weak systems from weak systems. $BSCR \approx 3$ is used to differentiate weak systems from strong systems. However, to the best knowledge of the authors, there is no quantitative indices as well as physical mechanism for the small signal stability analysis of PED systems. Moreover, the previous researches on SCR mainly focuses on the static voltage stability of AC-DC systems and the small signal stability of single power electronic based devices infeed systems (SPEIS). How to measure the strength and stability of MPEIS from the small signal stability point of view is still under challenge.

Compared with a single grid-connected device system, there exists complex interaction mechanisms not only between grid-connected devices and AC grid but also among grid-connected devices in multi grid-connected devices system, which increase the difficulty to analyze it. Fortunately, since the grid-connected devices are highly similar with multi-infeed system, and analytical method can be applicable for stability analysis. Hence, the concept of multi-infeed short circuit ratio is proposed aiming at simplifying multi-infeed system into single-infeed systems and extending the concept of SCR to analyze the multi-infeed systems. For example, *CIGRÉ* proposed multi-infeed short circuit ratio (MISCR) for multi-infeed HVDC systems (MIDC) by considering the neighboring HVDC's voltage influence [13]. However, MISCR is a rule-of-thumb extension of SCR and it is lacking of strict theory basis [14]. Ref. [15] proposed generalized short circuit ratio (gSCR) by eigenvalue decomposition from the voltage stability point of view, which overcome the rule-of-thumb basis of MISCR. Nevertheless, since the small signal stability problems are more likely to occur in typical MPEISs such as wind plants and photovoltaic plants [8,16], the gSCR index proposed in [15] is still not able to evaluate the stability of MPEIS.

In this paper, a static index named generalized short circuit ratio (gSCR) is proposed for the small signal stability evaluation of MPEIS. Firstly, the mathematical model of SPEIS is formulated via Jacobian transfer matrix. The relationship be-tween the SCR and small signal stability criterion is particularly analyzed. Then, based on the dynamic characteristics of multi infeed system, the MPEIS is decomposed into $n$ independent single infeed systems. In this regard, the concept of gSCR is proposed to analyze the small signal stability of MPEIS, which can be used to measure the strength of AC systems. Finally, derived from gSCR, OgSCR is proposed for multi VSCs infeed system. The analysis on participation factors and sensitivity to OgSCR and gSCR is conducted as well. The effectiveness of the proposed index is validated by simulations on the Matlab/Simulink.

## II. SINGLE-INFEED SHORT CIRCUIT RATIO

The typical SPEIS is shown as Fig. 1, which consists of two parts, i.e., the PED and the AC grid. $P$ and $Q$ are the real and reactive power output of PED. $U$ and $\theta$ are the magnitude and phase of PED terminal voltage. $L_g$ is the Thévenin inductance of AC grid.

### A. Characteristic equation of SPEIS

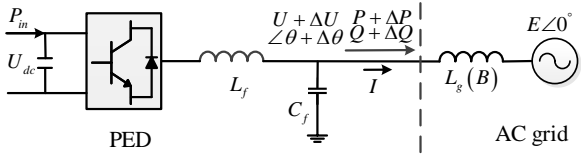

Fig. 1 Single power electronic based devices infeed systems

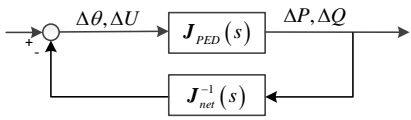

Fig. 2 Equivalent feedback-control diagram

The linearized input-output characteristics of PED and AC grid can be expressed via Jacobian transfer matrixes [17-18], given as follows:

$$\begin{bmatrix}\Delta P\\\Delta Q\end{bmatrix}=\underbrace{S_B\begin{bmatrix}G_{P\theta}(s) & G_{PU}(s)\\G_{Q\theta}(s) & G_{QU}(s)\end{bmatrix}}_{J_{PED\_s}(s)}\begin{bmatrix}\Delta\theta\\\Delta U/U\end{bmatrix} \quad (1)$$

$$\begin{bmatrix}\Delta P\\\Delta Q\end{bmatrix}=-\underbrace{\begin{bmatrix}J_{P\theta}(s) & J_{PU}(s)\\J_{Q\theta}(s) & J_{QU}(s)\end{bmatrix}}_{J_{net\_s}(s)}\begin{bmatrix}\Delta\theta\\\Delta U/U\end{bmatrix} \quad (2)$$

where $S_B$ is the rated capacity of PED, subscript $s$ represents SPEIS. And the negative sign in (2) indicates that $J_{net\_s}(s)$ has the opposite power positive direction to $J_{PED\_s}(s)$. The transfer functions in $J_{net\_s}(s)$ are given as in (3), with detailed derivation process shown in Appendix A.

$$\begin{cases}J_{P\theta}(s)=\alpha(s)BU_0^2-Q_0\\J_{PU}(s)=\beta(s)BU_0^2+P_0\\J_{Q\theta}(s)=-\beta(s)BU_0^2+P_0\\J_{QU}(s)=\alpha(s)BU_0^2+Q_0\end{cases} \quad (3)$$

where $B=\omega_0 L_g$, $\alpha(s)=\dfrac{1}{(s/\omega_0)^2+1}$, $\beta(s)=\dfrac{s/\omega_0}{(s/\omega_0)^2+1}$,

$\omega_0=2\pi*50$ is the nominal angular frequency of AC system, the subscript 0 denotes the initial value of steady operating point and will be omitted for convenience.

It can be observed from (1) and (2) that the closed-loop system becomes a multi-input-multi-output (MIMO) feedback control system shown as Fig. 2 and the characteristic equation is

$$\det\left[I_2+J_{PED\_s}(s)J_{net\_s}^{-1}(s)\right]=0 \quad (4)$$

where $I_2$ is 2-dimension identity matrix. Substituting (1) and (2) in (4) yields the detailed characteristic equation

$$c(s)=\det\left(S_B\begin{bmatrix}G_{P\theta}(s) & G_{PU}(s)\\G_{Q\theta}(s) & G_{QU}(s)\end{bmatrix}\right.$$
$$\left.+\begin{bmatrix}-\alpha(s)BU^2+Q & -\beta(s)BU^2-P\\\beta(s)BU^2-P & -\alpha(s)BU^2-Q\end{bmatrix}\right)=0 \quad (5)$$

### B. Relationship of SCR and small signal stability

In SPEIS, the SCR is defined as

$$SCR=\dfrac{S_{ac}}{S_B}=\dfrac{U_t^2}{Z}\times\dfrac{1}{S_B}=\dfrac{1}{S_B Z} \quad (6)$$

where $S_{ac}$ is the capacity of AC short circuit, $Z=1/B$ is the reactance of AC grid and the resistance is neglected to simplify the analysis.

Combining (5) and (6), the closed characteristic equation (5) can be rewritten as the determinant of Jacobian transfer matrixes (7) and the explicit function of SCR (8):

$$c(s)=\det\left(\dfrac{1}{U}\begin{bmatrix}G_{P\theta}(s)+Q_b & G_{PU}(s)-P_b\\G_{Q\theta}(s)-P_b & G_{QU}(s)-Q_b\end{bmatrix}\right.$$
$$\left.+\begin{bmatrix}-\alpha(s)\cdot SCR & -\beta(s)\cdot SCR\\\beta(s)\cdot SCR & -\alpha(s)\cdot SCR\end{bmatrix}\right)=0 \quad (7)$$

$$SCR^2+a(s)SCR+b(s)=0 \quad (8)$$

where $P_b=P/S_B$ and $Q_b=Q/S_B$ are the power output based on the rated capacity of AC system. $a(s)$ and $b(s)$ are given by

$$\begin{cases}a(s)=\dfrac{1}{[\alpha^2(s)+\beta^2(s)]U^2}\{-\alpha(s)[G_{P\theta}(s)+G_{QU}(s)]\\\qquad\qquad+\beta(s)[G_{Q\theta}(s)-G_{PU}(s)]\}\\b(s)=\dfrac{1}{[\alpha^2(s)+\beta^2(s)]U^4}\{G_{P\theta}(s)G_{QU}(s)-G_{PU}(s)G_{Q\theta}(s)\\\qquad\qquad+[G_{QU}(s)-G_{P\theta}(s)]Q_b+[G_{PU}(s)+G_{Q\theta}(s)]P_b-P_b^2-Q_b^2\}\end{cases}$$

Equation (8) reflects the explicit relationship between the small interference stability of the system and the short circuit ratio. Since the stability of the SPEIS increases with the increase of SCR [8,16,19], the small signal stability of SPEIS can be quantitatively evaluated via SCR.

**Definition1:** When SPEIS has two conjugate eigenvalues locating at the imaginary axis, the SCR is defined as the critical short circuit ratio (CSCR).

If SCR is smaller than CSCR, SPEIS is unstable and vice versa. Besides, the stability margin can be determined by the difference between SCR and CSCR.

Note that the PEDs are supposed to have a better dynamic property in a stronger system in the analysis above, which is satisfied by most kinds of PEDs such as the power electronic devices based on PLL and vector control. Since the detailed model and parameters of PED is not used in the derivation (4) -(8), the analysis above has generality. However, despite SCR is only related to the impedance of AC grid, CSCR and the relationship between stability margin and SCR is determined by the detail model of PED shown as (8).

## III. MULTI-INFEED SHORT CIRCUIT RATIO

The equivalent circuit of MPEIS is shown as Fig. 3. To simplify the stability analysis, the resistance is neglected, and the following assumptions hold in this paper:

**Assumption1:** The PEDs are similar, which means that the control strategy, the parameters and the compensation capacitors of PEDs based on their individual rated capacity are same.

**Assumption2:** The topology of AC system is connected and inductive. The nodal admittance matrix is reversible and symmetrical.

**Assumption3:** The power across interconnection lines is much less than its limitation.

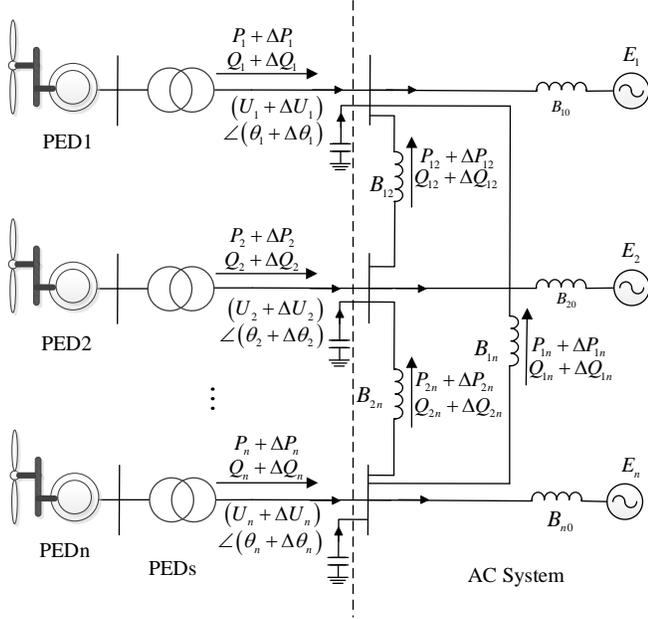

Fig. 3 Multi power electronic based devices infeed system

### A. Characteristic equation of MPEIS

Similar to SPEIS, MPEIS can be also regarded as a combination of PEDs and AC system, with linearized input-output characteristics expressed via Jacobian transfer matrixes $J_{PED\_m}(s)$ and $J_{net\_m}(s)$ respectively:

$$\begin{bmatrix} \Delta P \\ \Delta Q \end{bmatrix} = \underbrace{\left( \begin{bmatrix} G_{P\theta}(s) & G_{PU}(s) \\ G_{Q\theta}(s) & G_{QU}(s) \end{bmatrix} \otimes S_B \right)}_{J_{PED\_m}(s)} \begin{bmatrix} \Delta\theta \\ \Delta U/U \end{bmatrix}$$

$$\begin{bmatrix} \Delta P \\ \Delta Q \end{bmatrix} = -\underbrace{\begin{bmatrix} J_{P\theta}(s) & J_{PU}(s) \\ J_{Q\theta}(s) & J_{QU}(s) \end{bmatrix}}_{J_{net\_m}(s)} \begin{bmatrix} \Delta\theta \\ \Delta U/U \end{bmatrix} \quad (9)$$

where $\otimes$ denotes Kronecker product, subscript "$m$" denotes MEPIS, $S_B = diag(S_{Bi})$. And in the following article, $diag(a_i)$ represents diagonal matrix $diag(a_1, a_2...a_n)$ for convenience. The transfer functions in $J_{net\_m}(s)$ are given by (10), and the detailed derivation process is shown in Appendix A.

$$\begin{cases} J_{P\theta} = -\alpha(s)M - \beta(s)N - diag(Q_i) \\ J_{PU} = -\beta(s)M + \alpha(s)N + diag(P_i) \\ J_{Q\theta} = \beta(s)M - \alpha(s)N + diag(P_i) \\ J_{QU} = -\alpha(s)M - \beta(s)N + diag(Q_i) \end{cases} \quad (10)$$

where

$$\begin{cases} M_{ij} = U_i U_j B_{ij} \cos\theta_{ij} \\ N_{ij} = U_i U_j B_{ij} \sin\theta_{ij} \end{cases} \quad (11)$$

here $B_{ij}$ is the elements of node admittance matrix $B$, $B_{ii} > 0$.

MPEIS can be also regarded as $2n$-dimension MIMO feedback control system shown as Fig. 2 and the characteristic equation is

$$\det\left[ I_n + J_{PED\_m}(s) J_{net\_m}^{-1}(s) \right] = 0 \quad (12)$$

Since assumption 3 yields $|\sin(\theta_i - \theta_j)| \ll 1$, $N \approx 0$ and $U_i U_j \cos\theta_{ij} \approx U_i^2 \approx 1$. Equation (12) can be further simplified as

$$\det\left( \begin{bmatrix} G_{P\theta}(s) & G_{PU}(s) \\ G_{Q\theta}(s) & G_{QU}(s) \end{bmatrix} \otimes S_B + \begin{bmatrix} diag(Q_i) & -diag(P_i) \\ -diag(P_i) & -diag(Q_i) \end{bmatrix} + \begin{bmatrix} -\alpha(s)B & -\beta(s)B \\ \beta(s)B & -\alpha(s)B \end{bmatrix} \right) = \det(J_{sim}) = 0 \quad (13)$$

Note that the terminal voltage of PEDs $U_i$ is assumed to be 1p.u. in (13). However, this assumption is not necessary and the stability of system when PEDs not at rated operation point will be analyzed later in this paper via operation gSCR.

### B. MPEIS Decoupling

Multiplying (13) left by $I_2 \otimes S_B^{-1}$, (13) is equivalent to

$$\det\left( \begin{bmatrix} G_{P\theta}(s) & G_{PU}(s) \\ G_{Q\theta}(s) & G_{QU}(s) \end{bmatrix} \otimes I_n + \begin{bmatrix} -\alpha(s)J_{eq} + diag(Q_b) & -\beta(s)J_{eq} - diag(P_b) \\ \beta(s)J_{eq} - diag(P_b) & -\alpha(s)J_{eq} - diag(Q_b) \end{bmatrix} \right) = 0 \quad (14)$$

where $I_n$ is $n$-dimension identity matrix, $P_b = P_i/S_{Bi}$, $Q_b = Q_i/S_{Bi}$, $J_{eq}$ is defined as the extended Jacobian matrix

$$J_{eq} = S_B^{-1} B \quad (15)$$

According to assumption 2, all eigenvalues of the matrix $J_{eq}$ are positive. And the minimum eigenvalue of $J_{eq}$ is a simple eigenvalue, which means its geometric multiplicity and algebraic multiplicity is one, and the corresponding eigenvectors are positive as well [15]. Thus, there exists a matrix $T$ which decomposes the matrix $J_{eq}$ to diagonal matrix sorted as $0 < \lambda_1 < \lambda_2 \le \cdots \le \lambda_n$ by ascending order:

$$T^{-1} J_{eq} T = \Lambda = diag(\lambda_i) \quad (16)$$

Combining (15) and (16) yields

$$\det\left( \begin{bmatrix} G_{P\theta}(s) & G_{PU}(s) \\ G_{Q\theta}(s) & G_{QU}(s) \end{bmatrix} \otimes I_n + \begin{bmatrix} -\alpha(s)\Lambda - diag(Q_b) & -\beta(s)\Lambda + diag(P_b) \\ \beta(s)\Lambda + diag(P_b) & -\alpha(s)\Lambda + diag(Q_b) \end{bmatrix} \right) = 0 \quad (17)$$

Since all the elements in (17) are identical diagonal matrixes, we can rewrite (17) as follows

$$c_1(s) \times c_2(s) \cdots \times c_n(s) = 0 \quad (18)$$

where

$$c_i(s) = \det\left(\begin{bmatrix} G_{P\theta}(s)+Q_b & G_{PU}(s)-P_b \\ G_{Q\theta}(s)-P_b & G_{QU}(s)-Q_b \end{bmatrix} + \begin{bmatrix} -\alpha(s)\lambda_i & -\beta(s)\lambda_i \\ \beta(s)\lambda_i & -\alpha(s)\lambda_i \end{bmatrix}\right) = 0$$

It can be observed from (5) and (18) that the factors $c_i(s)$ in (18) is the same as $c(s)$ in (5), which means the closed characteristic equation of $n$-infeed system can be regarded as a production of $n$ independent SPEISs. The AC system rated capacity of equivalent SPEISs and MPEIS are assumed to be same for convenience, then the equivalent SPEISs follows the following equation

$$(P_{bi\_e}, Q_{bi\_e}, U_{i\_e}, B_{i\_e}) = (P_b, Q_b, 1, S_{Bi}\lambda_i) \quad (19)$$

where the subscript "$e$" donates equivalent single-infeed system. The SCR of each equivalent SPEIS is

$$SCR_i = S_{Bi}^{-1} \times S_{Bi}\lambda_i = \lambda_i \quad (20)$$

As analyzed above, the dynamic characteristics of MPEIS with $n$ PEDs can be represented by $n$ independent SPEISs, namely the $n$-infeed system can be decomposed into $n$ independent single infeed systems and the eigenvalues of MPEIS can be obtained by calculating the eigenvalues of $n$ equivalent SPEIS. Moreover, it can be observed from (19) that the equivalent SPEISs have the same parameters and operation condition except AC grid admittance. Therefore, the stability analysis of MPEIS can be transformed to the stability analysis of $n$ identical PEDs connected to different ac grid with different SCR.

*C. Generalized short circuit ratio (gSCR)*

Since the stability of MPEIS can be obtain by analyzing equivalent SPEISs, the necessary and sufficient conditions for the MPEIS to be stable is that all the equivalent SPEISs are stable. Therefore, the small signal stability of MPEIS is up to the weakest equivalent SPEIS, namely the SIPES corresponding to the smallest eigenvalue of $J_{eq}$. If the equivalent SIPES with the smallest SCR is stable, the MPEIS is stable and vice versa. Therefore, the gSCR for MPEIS is defined as follows.

**Definition2:** The minimum eigenvalue of extended Jacobian matrix $J_{eq}$ is defined as the generalized short circuit ratio, as

$$gSCR = \min \lambda(J_{eq}) \quad (21)$$

It can be seen from aforementioned analysis that SPEIS is a special MPEIS which dimension is 1 and the $J_{eq}$ of SPEIS is $\left[S_1^{-1}B_{11}\right]$. Hence, the properties of SCR can be extended to gSCR:

1) The gSCR is called critical gSCR (CgSCR) when MPEIS has two conjugate eigenvalues locating at the imaginary axis, which is used to differentiate unstable systems from stable systems.

2) The strength of AC grid and the stability of MPEIS increase with gSCR.

3) If gSCR is less than CgSCR, MPEIS is unstable, and vice versa. The stability margin can be evaluated by the difference between gSCR and CgSCR.

4) CgSCR of MPEIS equals the CSCR of the equivalent SPEISs numerically. Moreover, the relationship between gSCR and stability margin of MPEIS is the same as that of SPEIS.

There are two characteristics of gSCR. One the hand, gSCR is easy to calculate since it relates only to the node admittance matrix $B$ and rated capacity of the PEDs. One the other hand, although gSCR is a static index, it is able to evaluate the small signal stability of MPEIS. In other words, the dynamic part of AC grid ($\alpha(s)$ and $\beta(s)$) and PEDs are viewed as a whole shown as the coefficient $a(s)$ and $b(s)$ in (8), after MPEIS decomposition. Thus, gSCR is only determined by the static part of AC grid.

It is noticeable that the proposed gSCR index for MPEIS in this paper has the same expression as that for multi-infeed MIDC in [15]. And the gSCR for MIDC is a special case of gSCR for MPEIS from the eigenvalue point of view when $s=0$. However, these two gSCR indexes aim at different issues with different physical mechanism. The gSCR for MIDC is a static index aiming at evaluate the static stability of system, which defined from the static voltage stability point of view. And the gSCR for MPEIS is a static index aiming at evaluate the dynamic stability of system, which defined from the small signal stability point of view.

*D. Verification of Simplification in MPEIS Decoupling*

In this subsection, the influence of neglecting the elements related to $N$ in the simplification from (12) to (13) on eigenvalue calculation is discussed. The neglected part is given by

$$N_{neg} = \begin{bmatrix} -\beta(s)N & \alpha(s)N \\ -\alpha(s)N & -\beta(s)N \end{bmatrix} \quad (22)$$

Since the simplified MPEIS described by (13) is decomposed into $n$ equivalent SPEISs, the simplified MPEIS totally have $n(m+2)$ eigenvalues, where $m$ and $m+2$ is the number of state variables in a PED and a SPEIS respectively. Compared with $n$ equivalent SPEISs, the original MPEIS described by (12) has the transmission lines between the terminal of each PED. Thus, the original MPEIS has $n(m+2)+n(n-1)$ eigenvalues, where $n(n-1)$ is the number of state variables transmission lines between each PED. That is to say, the simplified MPEIS loses $n(n-1)$ eigenvalues after neglecting $N_{neg}$. However, $N_{neg}$ are mainly related to the dynamic of inductances. Thus, the lost eigenvalues are mainly determined by the dynamic of inductances, which have little influence on the stability analysis of MPEIS.

On the other hand, since the diagonal elements of $N_{neg}$ are all 0 and $|\sin\theta_{ij}| \ll 1$, the following equation is satisfied

$$\left\|N_{neg}\right|_{s=s_i}\right\|_F \ll \left\|J_{sim}\right|_{s=s_i}\right\|_F \quad (23)$$

where $s_i$ is the eigenvalues of the simplified MPEIS. It can be observed from (23) that the Frobenius norm of $N_{neg}$ is much less than that of $J_{sim}$, which means $N_{neg}$ has little influence on the remaining eigenvalues of original MPEIS. In summary, the influence of neglecting $N_{neg}$ on stability analysis is negligible.

IV. OPERATION GSCR FOR GRID-CONNNECTED VSC

As aforementioned discussion on gSCR, the power injection and terminal voltage of PEDs are assumed to be the same based on their individual capacity, which may not resemble the practical system. In this section, the operation gSCR (OgSCR) is deduced for the grid-connected VSCs to measure the strength of multi VSCs infeed power systems (MVIS) when the VSCs operate at an arbitrary operation point.

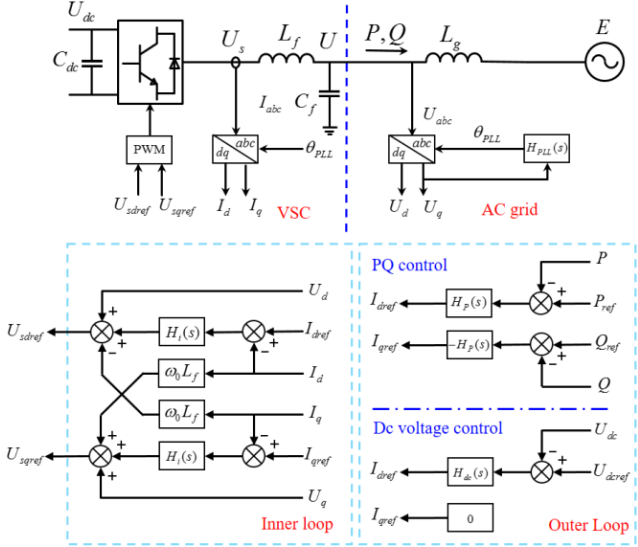

Fig. 4 Block diagram of the a single VSC infeed system

### A. Operation gSCR for Grid-connected VSCs

The VSC control system used in this paper can be referred to [20], which is designed as a PLL based dual-loop vector controller as shown in Fig. 4. The outer control loop can be designed as either output power controller or the dc voltage controller. Besides, the power factor of the converter is assumed to be closed to 1.

Based on the small signal model that developed in [20], the Jacobian transfer matrixes of two different types of VSCs can be derived and they all have the form as (24). The derivation process and the expression of $J_{VSC}(s)$ are shown in Appendix B

$$J_{PED\_s}(s) = S_B J_{VSC}(s) = S_B P_b \begin{bmatrix} UG_{oP\theta}(s) & UG_{oPU}(s)+1 \\ UG_{oQ\theta}(s)+1 & UG_{oQU}(s) \end{bmatrix} \quad (24)$$

where $P_b = P_i/S_B$ is the power injection based on the rated capacity of VSC.

Combining (9), (24) and considering assumption 3, the closed characteristic equation of MVIS can be expressed as:

$$\det\left(\begin{bmatrix} G_{oP\theta}(s) & G_{oPU}(s) \\ G_{oQ\theta}(s) & G_{oQU}(s) \end{bmatrix} \otimes (diag(U_i)P_b S_B) + \begin{bmatrix} & 1 \\ 1 & \end{bmatrix} \otimes (P_b S_B)\right.$$
$$\left. + \begin{bmatrix} -\alpha(s)M & -\beta(s)M \\ \beta(s)M & -\alpha(s)M \end{bmatrix} + \begin{bmatrix} diag(Q_i) & -diag(P_i) \\ -diag(P_i) & -diag(Q_i) \end{bmatrix}\right) = 0 \quad (25)$$

where $P_b = diag(P_{bi})$, $P_b = diag(P_i)S_B$ and $diag(Q_i) = 0$. Multiplying (25) left by $I_2 \otimes P_b^{-1}S_B^{-1}$ and right by $I_2 \otimes diag(U_i^{-1})$ yields

$$\det\left(\begin{bmatrix} G_{oP\theta}(s) & G_{oPU}(s) \\ G_{oQ\theta}(s) & G_{oQU}(s) \end{bmatrix} \otimes I_n + \begin{bmatrix} -\alpha(s)J_{eqo} & -\beta(s)J_{eqo} \\ \beta(s)J_{eqo} & -\alpha(s)J_{eqo} \end{bmatrix}\right) = 0 \quad (26)$$

where

$$J_{eqo} = P_b^{-1} S_B^{-1} M diag(U_i^{-1}) \approx diag(P_{bi}^{-1} S_{Bi}^{-1} U_i) B = diag(P_i^{-1} U_i) B \quad (27)$$

Similar to the derivation (16)-(18), there exists a matrix $T_o$ which decomposes the $J_{eqo}$ sorted as $0 < \lambda_{o1} < \lambda_{o2} \leq \cdots \leq \lambda_{on}$ by ascending order. Then (26) is simplified as

$$c_{o1}(s) \times c_{o2}(s) \cdots \times c_{on}(s) = 0 \quad (28)$$

where

$$c_{oi}(s) := \det\left(\begin{bmatrix} G_{P\theta}(s) & G_{PU}(s) \\ G_{Q\theta}(s) & G_{QU}(s) \end{bmatrix} + \begin{bmatrix} -\alpha(s)\lambda_{oi} & -\beta(s)\lambda_{oi} \\ \beta(s)\lambda_{oi} & -\alpha(s)\lambda_{oi} \end{bmatrix}\right) = 0$$

It can be observed from (28) that the MVIS can be decomposed into *n* equivalent independent single VSC infeed power systems (SVIS) which characteristic equations are $c_{oi}(s)$. Assuming the equivalent single-infeed systems are all working at rated operation point for convenience, then the parameters of SVIPS satisfies the following equation:

$$(P_{bi\_e}, Q_{bi\_e}, U_{i\_e}, S_{Bi\_e}, B_{i\_e}, SCR_i) = (1, 0, 1, 1, \lambda_{oi}, \lambda_{oi}) \quad (29)$$

Similar to the analysis of gSCR, the small signal stability of MVIS is up to the weakest equivalent SVIS, namely the SVIS corresponding to $\lambda_{oi}$. And the operation gSCR (OgSCR) is defined as follows:

**Definition3:** The minimum eigenvalue of $J_{eqo}$ is defined as the operation generalized short circuit ratio, as

$$OgSCR = \min \lambda(J_{eqo}) \quad (30)$$

It can be seen from (15) and (27) that $J_{eqo} = diag(P_{bi}^{-1} U_i) J_{eq}$, which means that gSCR is a special case of OgSCR when $diag(P_{bi}^{-1} U_i) = I$. In terms of the physical meaning, OgSCR is the extension of gSCR by taking the output power and terminal voltage into consideration. The OgSCR can be usedto reflect the influence of operation conditions of PEDs on the strength and stability of system. Besides, it is noticeable that OgSCR is no longer applicable when the Jacobian transfer matrix of PED does not meet the form as (24). However, gSCR does not have this restriction.

The proposed OgSCR has the following properties on the analogy of gSCR:

1) Critical OgSCR (COgSCR) is defined on the condition when MVIS has two conjugate eigenvalues locating at the imaginary axis. MVIS is unstable when OgSCR is less than COgSCR, and vice versa. Besides, the difference between gSCR and CgSCR represents stability margin.

2) COgSCR equals the CSCR. The relationship between OgSCR and stability margin of MVIS is the same as that of SVIS.

### B. Participation factors and sensitivity analysis

Based on the aforementioned analysis of OgSCR, MVIS can be decomposed into *n* equivalent SVIS, and each SVIS is a combination of the dynamic each PED on different scales. The participation factor of the *n*th PED to the *m*th equivalent SVISs can be indicated by

$$p_{om,n} = v_{omn} u_{onm} \quad (31)$$

where $v_o$ and $u_o$ are left and right eigenvector of $J_{eqo}$ respectively. Since $B$ is a symmetric matrix, the equation $diag(P_i^{-1} U_i) v_o^T = u_o$ (31) is equal to

$$p_{m,n} = v_{omn} u_{onm} = P_m^{-1} U_{om} u_{onm}^2 \quad (32)$$

Since the first equivalent SVIS, namely the SVIS corresponding to $\lambda_1$, has the weakest eigenvalue of MVIS, the participation factors of each VSC to SVIS1 reflects the participation degree of weakest eigenvalue of MVIS, which can be used to identify strongly correlated generators of weak damped modes.

Moreover, the sensitivity of rate capacity, output power and system admittance to OgSCR can be written as

$$\frac{\partial OgSCR}{\partial S_{Bi}} = \mathbf{v}_o \frac{\partial \mathbf{J}_{eqo}}{\partial S_{Bi}} \mathbf{u}_o = \mathbf{v}_o \frac{\partial diag\left(P_{bi}^{-1} S_{Bi}^{-1} U_i\right)}{\partial S_{Bi}} \mathbf{B} \mathbf{u}_o \quad (33)$$
$$= -\frac{P_{bi}}{U_i} OgSCR u_{oi}^2 < 0$$

$$\frac{\partial OgSCR}{\partial P_{bi}} = \mathbf{v}_o \frac{\partial \mathbf{J}_{eqo}}{\partial P_{bi}} \mathbf{u}_o = -\frac{S_{Bi}}{U_i} OgSCR u_{oi}^2 < 0 \quad (34)$$

$$\frac{\partial OgSCR}{\partial B_{ij}} = \mathbf{v}_o \frac{\partial diag\left(P_i^{-1} U_i\right)\mathbf{B}}{\partial B_{ij}} \mathbf{u}_o = \mathbf{u}_o^T \frac{\partial \mathbf{B}}{\partial B_{ij}} \mathbf{u}_o$$
$$= \begin{cases} (u_{o1i} - u_{o1j})^2 \geq 0 & i \neq j \cap j \neq 0 \\ u_{o1i} u_{o1j} > 0 & i = j \cup j = 0 \end{cases} \quad (35)$$

Sensitivity indicates the direction of improving system small signal stability and we can draw conclusions that: increasing system admittance, decreasing the capacity and the output power of PED can increase OgSCR, namely, improve system voltage stability.

Moreover, it can be seen from that (32)-(35) that participation factors and sensitivity are easy to calculate, since their expressions are only related to output power, admittance matrix, rated capacity and the right eigenvectors of $\mathbf{J}_{eqo}$.

Similarly, the participation factors and sensitivity expressions of gSCR can be derived as,

$$p_{m,n} = v_{mn} u_{nm} = S_{Bi} u_{nm}^2 \quad (36)$$

$$\frac{\partial gSCR}{\partial S_{Bi}} = -gSCR u_i^2 < 0 \quad (37)$$

$$\frac{\partial gSCR}{\partial B_{ij}} = \begin{cases} (u_{1i} - u_{1j})^2 \geq 0 & i \neq j \cap j \neq 0 \\ u_{1i} u_{1j} > 0 & i = j \cup j = 0 \end{cases} \quad (38)$$

where $v$ and $u$ are left and right eigenvector of $\mathbf{J}_{eq}$. The conclusions drawn from the sensitivity expressions is similar as that from OgSCR, namely, increasing system admittance and decreasing capacity of PED can improve system voltage stability.

## V. SIMULATION VALIDATION

In this section, case studies are carried out on a single VSC infeed system and a five VSCs infeed system to verify the effectiveness of multi-infeed system decoupling as well as the proposed OgSCR index. Since the gSCR is a special case of OgSCR, the verification of OgSCR also validates the effectiveness of gSCR. The PED model used in simulation is the VSC with dc-voltage control shown as Fig. 4 and the parameters are listed in Table 1.

### A. Analysis of SCR and CSCR

By changing the admittance of AC grid and solving characteristic equation (8), the relationship between SCR and the damping ratio of SVIS is obtained and plotted in Fig. 5.

Table 1. Parameters of VSC

| Symbol | Description | Value |
|---|---|---|
| $S$ | Base value of AC system | 1500kVA |
| $U_B$ | Base value of voltage | 690V |
| $L_f, C_f$ | Inductance and capacitance of convert | 0.05pu, 0.05pu |
| $H_{dc}(s)$ | Transfer function of the dc voltage controller | 0.2+200/s |
| $H_i(s)$ | Transfer function of the current controller | 0.6+15/s |
| $H_{pll}(s)$ | Transfer function of the PLL | 2+3020/s |

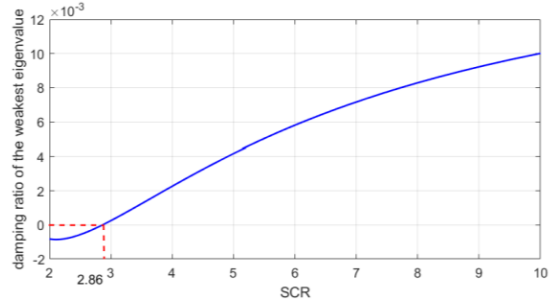

Fig. 5 The Relationship between Stability Margin and SCR

It can be seen from Fig. 5 that the damping ratio increases with SCR and the SVIS has two conjugate imaginary eigenvalues when SCR=2.86, i.e., CSCR is 2.86. Therefore, the SVIS is stable only if SCR is larger than 2.86. Since there is a positive correlation between damping ratio of the weakest eigenvalue of SIVS and SCR, the difference between SCR and CSCR can evaluate the stability margin, which met the properties of SCR analyzed above.

### B. MVIS decoupling and OgSCR verification

The five VSCs infeed system is constructed with the parameters given in Table 2 and Table 3.

Table 2. Parameters of AC grid

| Admittance | Value(p.u.) | Admittance | Value(p.u.) | Admittance | Value(p.u.) |
|---|---|---|---|---|---|
| $B_{10}$ | 0.2 | $B_{20}$ | 0.15 | $B_{34}$ | 0.07 |
| $B_{12}$ | 0.15 | $B_{23}$ | 0.18 | $B_{35}$ | 0.05 |
| $B_{13}$ | 0.1 | $B_{24}$ | 0.2 | $B_{40}$ | 0.1 |
| $B_{14}$ | 0.06 | $B_{25}$ | 0.21 | $B_{45}$ | 0.11 |
| $B_{15}$ | 0.09 | $B_{30}$ | 0.25 | $B_{50}$ | 0.2 |

Table 3. Parameters of VSCs

| | VSC1 | VSC2 | VSC3 | VSC4 | VSC5 |
|---|---|---|---|---|---|
| $P_{bi}$ | 0.8 | 0.7 | 0.9 | 1 | 0.5 |
| $S_{Bi}$ | 1.5 | 2 | 1 | 1.8 | 1.5 |

Table 4. The relationship between equivalent SVISs and the eigenvalue of $\mathbf{J}_{eqo}$

| SVIS | 1 | 2 | 3 | 4 | 5 |
|---|---|---|---|---|---|
| $\lambda_i$ | 6.0944 | 24.9627 | 46.6669 | 56.6602 | 94.3062 |

Firstly, decomposing the five VSCs infeed system into five SIVSs, and the corresponding eigenvalue of $\mathbf{J}_{eqo}$ is given in Table 4. The eigenvalues directly calculated by five-infeed system and those of the five equivalent SVISs are shown in Fig. 6. It can be seen that the eigenvalues calculated by MVIS SVISs are almost the same. Therefore, the equivalent SVISs can exactly represent the stability of MVIS, which verifies the rationality of multi-infeed system decoupling.

Moreover, it can be seen from Table 4 shows that OgSCR is 6.0944 since the smallest eigenvalue of $\mathbf{J}_{eq}$ is $\lambda_1$. From Fig. 5 it can be observed that OgSCR satisfies OgSCR> COgSCR = CSCR=2.86 and the damping ratio of the weakest eigenvalue is 0.0059, which means MPEIS is stable. On the other hand, t the weakest eigenvalue is -0.3132 +53.3411i and the damping ratio is 0.0059, which is the same as the conclusion obtained by OgSCR. This means the stability assessment from gSCR is effective.

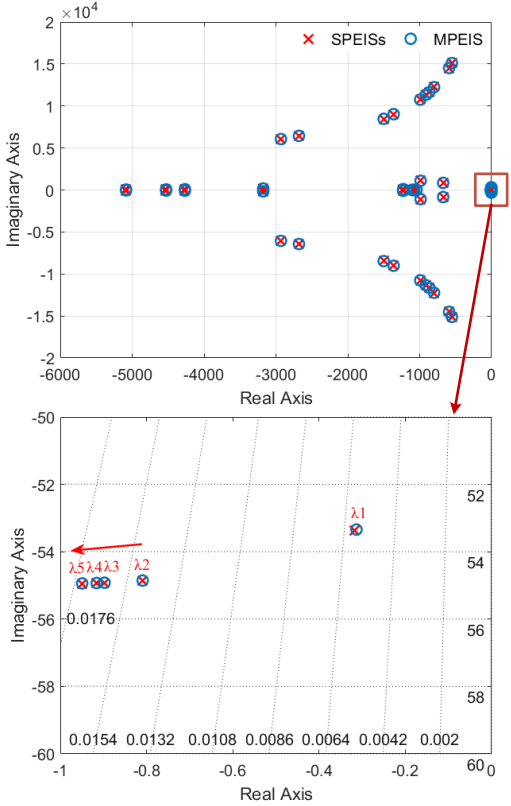

Fig. 6 Eigenvalues comparison between MEPIS and equivalent SIPESs

### C. Participation factors and sensitivity analysis

To verify the effectiveness of participation factors on strongly correlated generators identification, the participation factors of each VSC to the weakest equivalent SVIS and the participation factors of state variables in MVIS to the weakest eigenvalue (-0.3132 +53.3411) are calculated and given in Table 5.

Table 5. Participation factors to the weakest eigenvalue of MVIS

| VSC | Participation factors to weakest SVIS | State Variables | Participation factor to the weakest eigenvalue |
|---|---|---|---|
| 4 | 1 | PLL integrator | 1 |
|   |   | PLL PI | 0.9991 |
| 2 | 0.8730 | PLL integrator | 0.8689 |
|   |   | PLL PI | 0.8681 |
| 1 | 0.6948 | PLL integrator | 0.6935 |
|   |   | PLL PI | 0.6928 |
| 3 | 0.4810 | PLL integrator | 0.4810 |
|   |   | PLL PI | 0.4806 |
| 5 | 0.3029 | PLL integrator | 0.3028 |
|   |   | PLL PI | 0.3025 |

It can be seen from Table 5 that the participation factors of each VSC to the weakest SVIS and the weakest eigenvalue are almost the same, which means the participation factors of each VSC to the weakest SVIS can reflect the participation degrees of each VSC weakest eigenvalue and are able to be used for strongly correlated generators identification.

Moreover, Fig. 7 gives the trajectories of the OgSCR when the output power $P_{b1}$ of VSC1 or line admittance $B_{10}$ varies. It can be seen clearly that OgSCR decrease when PED output power increase or line admittance decrease, which meet the conclusion of sensitivity analysis.

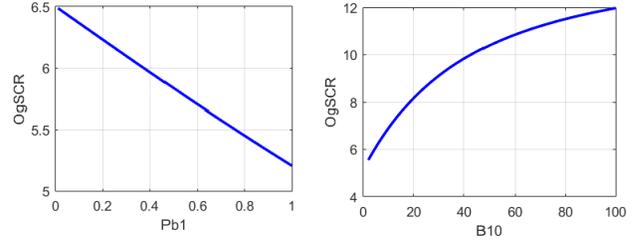

Fig. 7 OgSCR variation with power and admittance

## VI. CONCLUSION

Based on multi-infeed system decoupling method, the concept of gSCR is proposed to analyze the small signal stability of multi infeed system, which can be further used to judge the strength of AC systems. The gSCR is defined from the point of view of small signal stability, which physically and mathematically unify the concept of SCR in single and multi-infeed system. The proposed gSCR is a static index and easy to calculate using the proposed method, which provide theoretical support for the analysis of multi identical devices infeed system. Moreover, the OgSCR are proposed as the extension and application of gSCR. Simulation results validate the effectiveness of OgSCR and gSCR. Further work will be extended to exploring the interaction of nonidentical PEDs and the corresponding stability evaluation method.

## Appendix

### A. Derivation of $J_{net\_s}(s)$ and $J_{net\_m}(s)$

Since the AC grid in SPEIS is the special case of AC grid in MPEIS when $n=1$, we only derive $J_{net\_m}(s)$ here and $J_{net\_s}(s)$ equals to $J_{net\_m}(s)$ when $n=1$.

The dynamic equations of inductance between node $i$ and $j$ is shown as follows:

$$\begin{bmatrix} U_{ix} \\ U_{iy} \end{bmatrix} - \begin{bmatrix} U_{jx} \\ U_{jy} \end{bmatrix} = \begin{bmatrix} sL_{ij} & -\omega_0 L_{ij} \\ \omega_0 L_{ij} & sL_{ij} \end{bmatrix} \begin{bmatrix} I_{ijx} \\ I_{ijy} \end{bmatrix} \quad (39)$$

$$\begin{cases} P_{ij} = U_{ix} I_{ix} + U_{iy} I_{iy} \\ Q_{ij} = -U_{ix} I_{iy} + U_{iy} I_{ix} \end{cases} \quad (40)$$

$$\begin{bmatrix} P_i \\ Q_i \end{bmatrix} - \begin{bmatrix} \sum_{j=0, j\neq i} P_{ij} \\ \sum_{j=0, j\neq i} Q_{ij} \end{bmatrix} = \begin{bmatrix} 0 \\ 0 \end{bmatrix} \quad (41)$$

where $I$ is the output current of PED, subscribe "$x$" and "$y$" represent the $x$-axis component and $y$-axis component in synchronous rotating coordinate respectively.

The transformation of voltage form rectangular coordinate form into polar form is expressed as:

$$\begin{cases} U_{ix} = U_i \cos\theta_i \\ U_{iy} = U_i \sin\theta_i \end{cases} \quad (42)$$

In addition, the following equations are satisfied at steady

operating point:

$$\begin{cases} I_{ijd} = B_{ij}\left(U_{iq} - U_{jq}\right), I_{ijq} = B_{ij}\left(-U_{id} + U_{jd}\right) \\ P_i = \sum_{j=0}^{n} U_i U_j B_{ij} \sin\theta_{ij}, Q_i = \sum_{j=0}^{n} U_i U_j B_{ij} \cos\theta_{ij} \end{cases} \quad (43)$$

Linearizing (39)-(42), and combining them with (43) yields the Jacobian transfer matrixes $J_{net\_m}(s)$:

$$\begin{cases} J_{P\theta} = -\alpha(s)M - \beta(s)N - diag(Q_i) \\ J_{PU} = -\beta(s)M + \alpha(s)N + diag(P_i) \\ J_{Q\theta} = \beta(s)M - \alpha(s)N + diag(P_i) \\ J_{QU} = -\alpha(s)M - \beta(s)N + diag(Q_i) \end{cases} \quad (44)$$

B. *Derivation of Jacobian Transfer Matrix of VSC*

The admittance matrix $Y_{PQ}(s)$ and $Y_{dc}(s)$ of the converter with PQ control and dc-voltage control in polar coordinates are given as (45) and (46) respectively[20]:

$$\begin{bmatrix} \Delta I \\ I\Delta\varphi \end{bmatrix} = I \underbrace{\begin{bmatrix} Y_{PQ11}(s) & 0 \\ 0 & Y_{PQ22}(s) \end{bmatrix}}_{Y_{PQ}(s)} \begin{bmatrix} \Delta U \\ U\Delta\theta \end{bmatrix} \quad (45)$$

$$\begin{bmatrix} \Delta I \\ I\Delta\varphi \end{bmatrix} = I \underbrace{\begin{bmatrix} Y_{dc11}(s) & 0 \\ 0 & Y_{dc22}(s) \end{bmatrix}}_{Y_{dc}(s)} \begin{bmatrix} \Delta U \\ U\Delta\theta \end{bmatrix} \quad (46)$$

where $\varphi$ is the angle of output current

$$\begin{cases} Y_{PQ11} = -\dfrac{H_P H_i(s)}{H_P H_i U + H_i + sL_f} \\ Y_{PQ22} = \dfrac{H_P H_i - sL_f H_{pll}}{(H_P H_i U + H_i + sL_f)(1 + UH_{pll})} + \dfrac{H_{pll}}{1 + UH_{pll}} \end{cases}$$

$$\begin{cases} Y_{dc11} = -\dfrac{H_i H_{dc}}{U_{dc} Cs(sL_f + H_i) + H_i H_{dc} U_d} \\ Y_{dc22} = \dfrac{H_i H_{pll}}{(sL_f + H_i)(1 + U_d H_{pll})} \end{cases}$$

here $H_P = K_{pP} + K_{iP}/s$, $H_{dc} = K_{pdc} + K_{idc}/s$, $H_i = K_{pi} + K_{ii}/s$ and $H_{pll} = (K_{ppll} + K_{ipll}/s)/s$ are transfer functions of PQ control, dc voltage control, current control and PLL respectively. $L_f$ is the filter inductance.

The output power of VSC can be expressed

$$\begin{cases} P = UI\cos(\theta - \varphi) \\ Q = UI\sin(\theta - \varphi) \end{cases} \quad (47)$$

Linearizing (47) yields

$$\begin{bmatrix} \Delta P \\ \Delta Q \end{bmatrix} = \begin{bmatrix} P \\ & P \end{bmatrix} \begin{bmatrix} \Delta\theta \\ \Delta U/U \end{bmatrix} + \begin{bmatrix} & P \\ -P & \end{bmatrix} \begin{bmatrix} \Delta\varphi \\ \Delta I/I \end{bmatrix} \quad (48)$$

Substituting (45)-(46) into (48) yields Jacobian transfer matrix $J_{VSC\_PQ}(s)$ and $J_{VSC\_dc}(s)$ as (49) and

$$J_{VSC\_PQ}(s) = P\begin{bmatrix} & UY_{PQ11}(s)+1 \\ -UY_{PQ22}(s)+1 & \end{bmatrix} \quad (49)$$

$$J_{VSC\_dc}(s) = P\begin{bmatrix} & UY_{dc11}(s)+1 \\ -UY_{dc22}(s)+1 & \end{bmatrix} \quad (50)$$